\documentclass[twocolumn,aps,pra,amsmath,amssymb]{revtex4-1}

\usepackage{graphicx}
\usepackage{bm}
\usepackage{lipsum}
\usepackage{epstopdf}
\usepackage{epsfig}
\usepackage{dsfont}
\usepackage{amssymb}

\usepackage{braket}

\usepackage{leftidx}
\begin{document}

\begin{center}
{\Large Entanglement statistics of randomly interacting spins}

\vspace{1cm}

Paulo Freitas Gomes

Instituto de Ci\^encias Exatas e Tecnol\'ogicas, Universidade Federal de Jata\'{\i}, Jata\'{\i}, GO 75801-615, Brazil
\vspace{0.8cm}

Marcel Novaes

Instituto de F\'{\i}sica, Universidade Federal de Uberl\^andia, Uberl\^andia, MG 38408-100, Brazil
\vspace{0.8cm}

Fernando Parisio

Departamento de F\'{\i}sica, CCEN, Universidade Federal de Pernambuco, Recife, PE 50670-901, Brazil

\end{center}

\vspace{1cm}

{\bf We investigate the entanglement in the ground state of systems comprising two and three qubits with random interactions. Since the Hamiltonians also contain deterministic one-body terms, by varying the interaction strength, one can continuously interpolate between deterministic separable eigenstates and fully random entangled eigenstates, with non-trivial intermediate behavior. Entanglement strongly depends on the underlying topology of the interaction among the qubits. For a certain class of interactions GHZ entanglement is favoured by a non-separable collective interaction, while for fully separable pairwise interactions the ground states concentrate in the vicinity of W states.}

\section{Introduction}
\label{intro}
Randomness and entanglement are two fundamental features of quantum mechanics. While the former can ultimately be traced back to Born's postulate \cite{alexia}, the latter arises from the tensor structure of composite Hilbert spaces and the superposition principle \cite{horodecki}, in the standard framework of quantum theory. From the perspective of quantum information science, both are valuable resources for executing communication and information processing tasks.

Apart from randomness with purely quantum origins, stochasticity may appear as a consequence of the complexity of interactions among the several parts of a physical system, as, for instance, in atomic nuclei \cite{wigner}. This may be modeled by random Hamiltonians, which typically also generate entanglement. A direct approach to the problem is to bypass the analysis of Hamiltonians and study random states directly, by considering random rotations of a reference vector, for example. This sort of state has been studied analytically and numerically from various perspectives, mostly focusing on bipartite entanglement \cite{page,cappellini,giraud,marko,nadal1,nadal2,vivo}. Recently, the entanglement properties of three-qubit random states have also been studied \cite{enriquez}. 

However, this kind of {\it completely} random state does not naturally occur in all physical situations involving randomness. A perhaps more natural problem is to study of states associated with Hamiltonians which are partially random but retain some kind of structure, either through their eigenvectors or via dynamics. There have been many investigations in this direction, particularly in connection with the problem of thermalization \cite{pineda,arul,magan,alessio,wick,rigol,grover,vidmar}.

In this work, we study entanglement in ground states of low-dimensional qubit systems in which interactions are fully random, but the Hamiltonians also contain deterministic one-body terms. The situation can be thought of as a spin lattice with random interactions, where each spin is subjected to a constant magnetic field. The relative intensity between one-body terms and interaction can be varied continuously, tuning the eigenstates from deterministic to fully random, with interesting intermediate phenomena. 

We show that the extent and statistical properties of bipartite and tripartite entanglement heavily depends on the topological nature of the interaction potentials.

In the next section we provide the necessary background on random matrices and entanglement quantification. Sec. III addresses the simple case of two qubits. Sections IV and V consider systems comprising three qubits. We provide closing remarks in Sec. VI.

\section{Preliminary concepts}
\subsection{Random matrices}
Complex Hamiltonians have long been modeled as random matrices in a variety of systems, from nuclear physics to quantum dots, from microwave billiards to disordered media \cite{haake,handbook,applications}. The simplest model is to enforce hermiticity but draw the samples according to a Gaussian measure,
\begin{equation}
P({\cal H})\propto e^{-{\rm Tr}({\cal H}^2)/\sigma}.
\end{equation}
This has two important consequences: first, the matrix elements are independent, identically distributed random variables; second, the ensemble is rotation-invariant, in the sense that $P(U{\cal H}U^\dagger)=P({\cal H})$ for any unitary transformation $U$. In the absence of any specific symmetries, the matrix is complex hermitian without any further constraint. If it has dimension $N$, this is called the $GUE(N)$ -- Gaussian Unitary Ensemble \cite{Livan2018}.

Modeling the Hamiltonian as a random matrix means giving up the ambition of obtaining results that describe any specific system and, instead, focusing on properties that may be of universal validity, representative of systems that are typical in some sense, or set a null hypothesis against which any results or conjectures supposed to hold for a given system may be compared.   

The eigenstates of a matrix from $GUE(N)$ may be arranged as a unitary matrix, and this is distributed uniformly in the unitary group. We will refer to this by the standard terminology of {\it Haar measure}. Basically, normalized eigenstates are uniformly distributed as points in the complex sphere.

The parameter $\sigma$ controls the variance of the interactions, such that for small $\sigma$ the matrix ${\cal H}$ becomes vanishingly small, while for large $\sigma$ the matrix ${\cal H}$ may have very large elements, in the sense that
\begin{equation}
\langle {\cal H}_{ij}\rangle=0,\quad \langle |{\cal H}_{ij}|^2 \rangle \propto \sigma.
\end{equation}
For a single spin, for example, with the total Hamiltonian given by $\hat{\sigma}_z+V$, where $\hat{\sigma}_z$ is the usual Pauli matrix and $V$ is taken at random from the $GUE(2)$, the situation will be as follows: for small $\sigma$, the eigenvalues will be close to $\pm\frac{1}{2}$ and the eigenvectors will be close to $|0\rangle$ and  $|1\rangle$; for large $\sigma$, the eigenvalues will be a pair of correlated random variables and the eigenstates will be random vectors with Haar measure.
\subsection{Entanglement involving two and three qubits}
Entanglement \cite{horodecki} quantification is a difficult problem and, in general, the degree of non-separability embodied by a quantum state is not uniquely captured by a single figure of merit. Different quantifiers may lead to distinct orderings. However, when it comes to two or even three qubits, things simplify considerably.
In the first case we employ the concurrence \cite{wootters} as the entanglement quantifier, whereas in the second, more complex case we characterize non-separability with both the concurrence (of bipartitions and reduced states) and the three-tangle \cite{3tangle}.

Given an arbitrary two-qubit state represented by a density matrix $\rho$, its concurrence is determined by the eigenvalues $\left\lbrace \lambda_1, \lambda_2, \lambda_3, \lambda_4 \right\rbrace$, with $\lambda_1 \geq \lambda_2 \geq \lambda_3 \geq \lambda_4$, of the matrix $\rho \tilde{\rho}$, where $\tilde{\rho} = \left(  \sigma_y \otimes \sigma_y \right)  \rho^* \left(  \sigma_y \otimes \sigma_y \right)$, and given by 
\begin{equation*}
C={\rm max}\{\sqrt{\lambda_1} - \sqrt{\lambda_2} - \sqrt{\lambda_3} - \sqrt{\lambda_4},0\}. 
\end{equation*}
This reduces to $C=2 \vert ad-bc \vert$, for the general pure state $|\psi\rangle=a|00\rangle+b|01\rangle+c|10\rangle+d|11\rangle$. The squared concurrence is often referred to in the literature as the two-tangle.

For a pure state $\vert \Psi\rangle$ of three qubits, the three-tangle, or residual entanglement, $\tau$, is a quantifier of genuine three-partite entanglement, which is non-zero only for the class of Greenberger-Horne-Zeilinger (GHZ) states, vanishing for fully separable, bi-separable, and W states \cite{3qubits}. It is given by 
\begin{equation}
\tau=C^2_{1|23}-C^2_{12}-C^2_{13},
\label{tau}
\end{equation}
where $1,2,3$ are labels for the three qubits; ${1|23}$ refers to the bipartition where $1$ is a subsystem and $23$ is the other subsystem, while $12$ and $13$ denote reduced systems, where $3$ and $2$ have been traced out, respectively. The concurrence of $1|23$ is $C_{1|23}=2\sqrt{\det \rho_1}$ \cite{3tangle}, where $\rho_1={\rm Tr}_{23}(|\Psi\rangle\langle \Psi|)$. 

Although not evident from the definition, the three-tangle is invariant under any permutation of the subsystems, being a feature of the whole system. 

We will also refer to the total concurrence $C_t=C_{12}+C_{13}+C_{23}$. For any pure state of three qubits we have $C_t\le 4/3$ (the equality holds if and only if $|\Psi\rangle$ is a maximally entangled W state) \cite{3qubits}. 
\section{Two-qubit random eigenstates}
We begin our study by considering a four-dimensional Hilbert space, ${\cal H}={\cal H}_1\otimes {\cal H}_2$, with ${\cal H}_i=\mathds{C}^{\otimes 2}$ being the Hilbert space of a spin, the total Hamiltonian being given by:
\begin{equation}
H=H_1+H_2+V,
\end{equation}
with
\begin{equation}
H_1=\hat{\sigma}_{z,1}\otimes \mathds{1}_2,\;\; H_2=\mathds{1}_1\otimes \hat{\sigma}_{z,2},
\end{equation}
where $\hat{\sigma}_{z,i}$ stands for the Pauli operator in the $z$ direction, acting on ${\cal H}_i$. 

The interaction term $V$ is random in one of the two following ways. Either $V=V_{1}\otimes V_2$ (i), where $V_i$ are taken from $GUE(2)$, or else $V=V_{12}$ (ii) is directly taken from $GUE(4)$. 

To make meaningful comparisons between situations (i) and (ii), we must guarantee that the interaction term $V$ has the same average magnitude for a given $\sigma$. That is to say, when we take $\sigma_{12}=\sigma$ as the variance of the Gaussian ensemble related to $V_{12}$, we choose $\sigma_1=\sigma_2=\sqrt{\sigma}$ for $V_1$ and $V_2$, respectively. With this we get 
$$ \braket{ \vert (V_{12})_{ij}\vert ^2 } =  \langle | (V_{1}\otimes V_2)_{ij}|^2\rangle\propto \sigma.$$

Considering $\sigma_1 = \sigma_2$ incurs in no loss of generality, since one can easily show that the results depend only on the product $\sigma_1 \sigma_2$ (which equals $\sigma_{12}$). So, in the reminder of this section we will consider a single variance parameter $\sigma$, as described in the previous paragraph. Both kinds of interactions $V$ vanish when $\sigma = 0$, leading to a direct sum total Hamiltonian $H_1+H_2$ with a separable ground state.

\begin{figure}[h]
\includegraphics[height=5cm,angle=0]{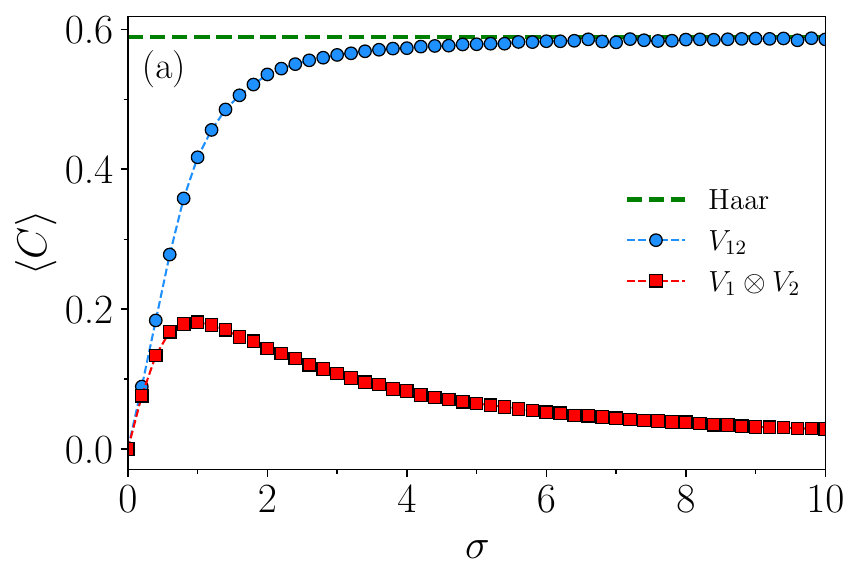}
\includegraphics[height=5cm,angle=0]{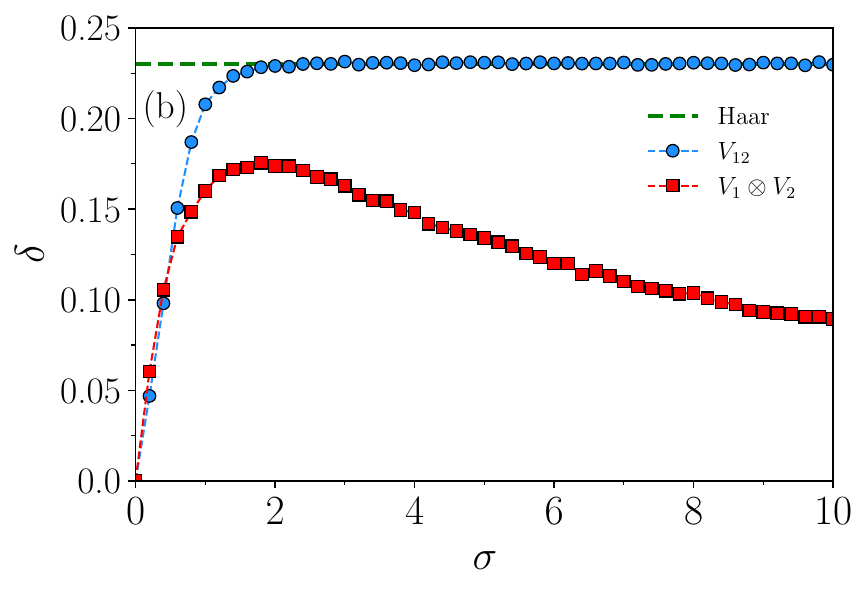}
\caption{(a) Average concurrence $\left\langle C \right\rangle $ as a function of $\sigma$ with $N=5 \times 10^4$ samples for the two types of interaction potential: $V_{12}$ (blue circles) and $V_1 \otimes V_2$ (red squares). (b) Standard deviations of $C$ as a function of $\sigma$. Horizontal lines indicate the results for Haar-random states. (color online)} 
\label{fig1} 
\end{figure}

In Fig. \ref{fig1}(a) we plot the average ground-state concurrence, computed from $5 \times 10^4$ realizations, as a function of the parameter $\sigma$. 

In the regime $\sigma \ll 1$, we found that $\langle C \rangle \sim \sigma$, for both kinds of interactions. This asymptotics can easily be derived from perturbation theory. More interestingly, in case (i) the average concurrence attains a maximum for $\sigma=1$ and vanishes asymptotically, whereas in situation (ii) it grows monotonically with $\sigma$, tending to the corresponding value for a Haar random state, which is given by \cite{Zyc} 
\begin{equation}\langle C \rangle_{Haar}=
\frac{3\pi}{16} \approx 0.589.
\end{equation} 
5
These two distinct behaviors for different kinds of interaction can be understood as follows. When $\sigma \rightarrow \infty$, the terms $H_1$ and $H_2$ in the Hamiltonian become negligible compared to $V$, i. e., $H \approx V $. In case (i) this leads to a Hamiltonian with separable eigenstates, whereas in case (ii) the Hamiltonian remains non-factorable. 

In Fig. \ref{fig1}(b)  we plot the concurrence standard deviation $\delta=\sqrt{\langle C^2 \rangle-\langle C \rangle^2}$ (in the current context of two qubits we can see that the curves $\langle C\rangle(\sigma)$ and $\delta(\sigma)$ are very similar, but this is not always the case for three qubits).

As expected, under interaction $V_{12}$ the quantity $\delta$ rapidly converges to the corresponding value of Haar random states (these have $\langle C^2 \rangle_{Haar}=2/5$, which leads to $\delta_{Haar}\approx 0.23$ \cite{Zyc,cappellini}).

We note that, whereas in case (ii) the average concurrence is about twice as large as the corresponding standard deviation, in case (i) both $\langle C\rangle$ and $\delta$ have similar magnitudes for a given $\sigma$. 
\section{Three qubits, collective interaction}

Now we investigate the degree of entanglement of three-qubit ground states, thus, in an eight-dimensional Hilbert space, ${\cal H}={\cal H}_1\otimes {\cal H}_2\otimes {\cal H}_3$, with ${\cal H}_i=\mathds{C}^{\otimes 2}$, $i=1,2,3$. The considered random Hamiltonians are given by
\begin{equation}
H=H_1+H_2+H_3+V,
\label{ham3}
\end{equation}
where
$H_1=\hat{\sigma}_{z,1} \otimes \mathds{1}_2 \otimes \mathds{1}_3$, $H_2=\mathds{1}_1 \otimes \hat{\sigma}_{z,2} \otimes \mathds{1}_3$, and 
$H_3=\mathds{1}_1\otimes \mathds{1}_2\otimes \hat{\sigma}_{z,3}$.
\begin{figure}[h]
\includegraphics[height=5cm,angle=0]{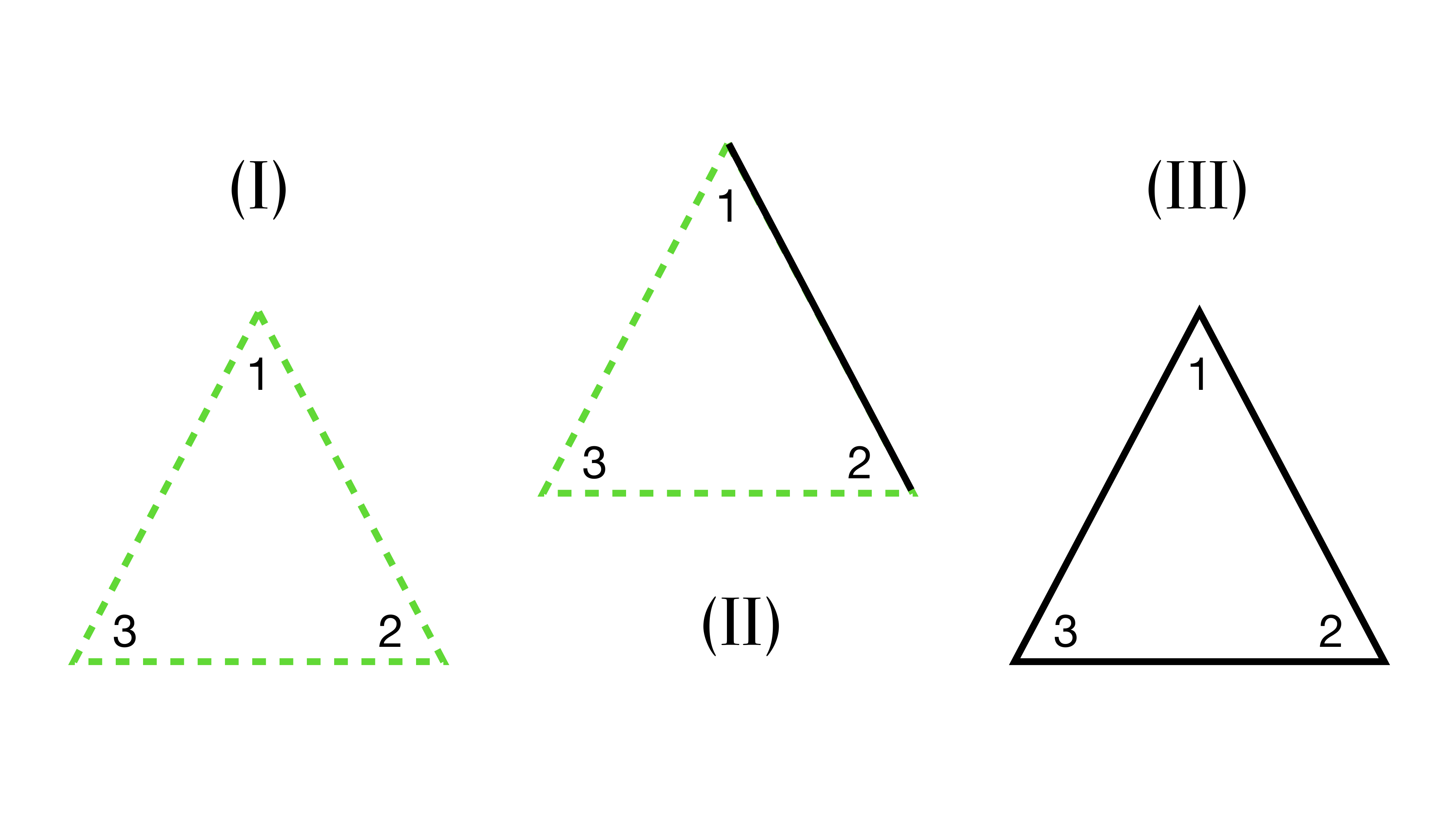}
\caption{Diagrams representing the three types of interactions considered in this section. In the triangles, qubits are the corners and interactions are the sides. Bold segments represent non-separable interactions and dashed segments represent separable interactions. (color online)} 
\label{fig2} 
\end{figure}

\begin{figure}[h]
\includegraphics[height=5cm,angle=0]{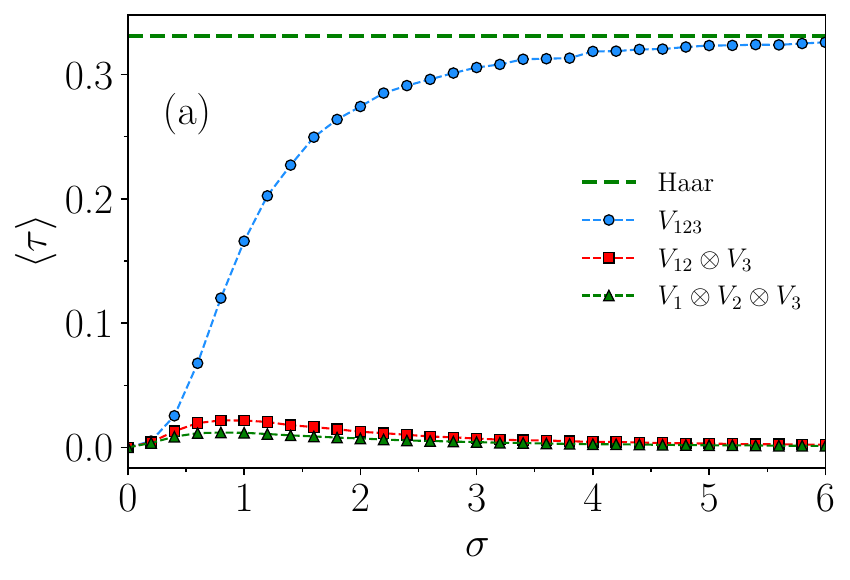}
\includegraphics[height=5cm,angle=0]{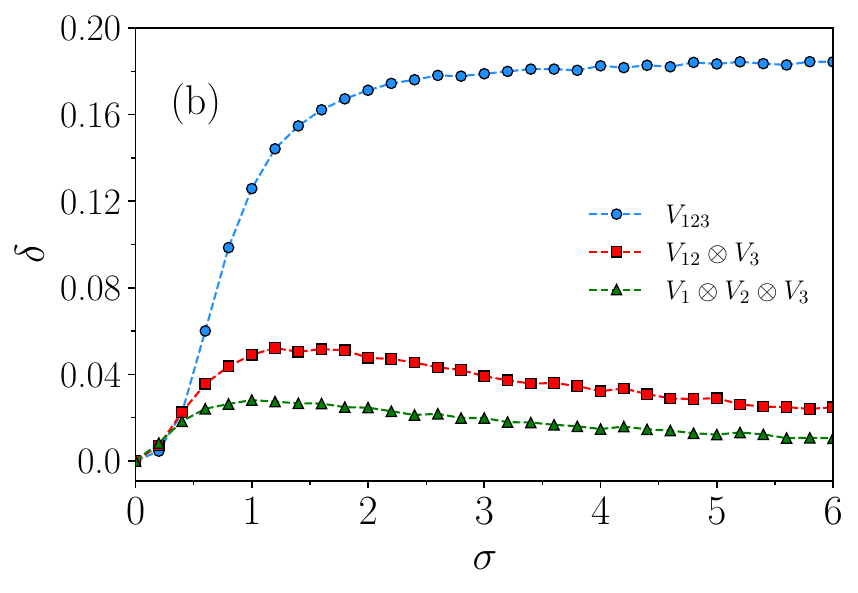}
\caption{(a) Average 3-tangle $\left\langle \tau \right\rangle $ with $5 \times 10^4$ samples for different interactions. (b) Corresponding standard deviation $\delta$. (color online)}
 \label{fig3} 
\end{figure}
\begin{figure}[h]
\includegraphics[height=5cm,angle=0]{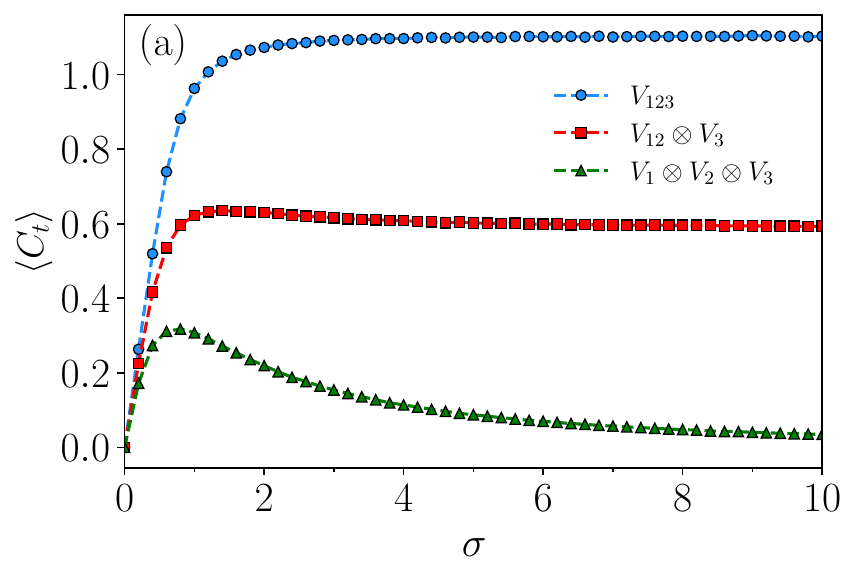}
\includegraphics[height=5cm,angle=0]{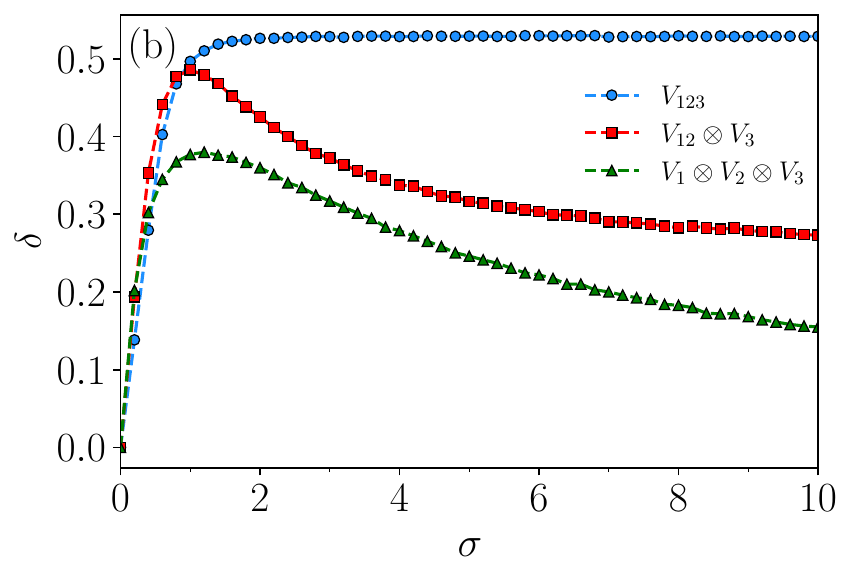}
\caption{(a) Total concurrence $\left\langle C_t \right\rangle $ with $\rho_{12}$ and $15 \times 10^4$ samples for different interactions. (b) Corresponding standard deviation $\delta$. (color online)} 
\label{fig4} 
\end{figure}
In this case, a larger variety of nonequivalent potentials exist. Initially we study the following interactions: 
\begin{equation} V_I=V_{1}\otimes V_2\otimes V_3,\end{equation} with each $V_i$ in $GUE(2)$, fully separable; 
\begin{equation} V_{II}=V_{12}\otimes V_3,\end{equation} where $V_{12}$ and $V_3$ are in $GUE(4)$ and $GUE(2)$, respectively, partially separable; and
\begin{equation} V_{III}=V_{123},\end{equation} in $GUE(8)$, fully non-separable. See Fig. \ref{fig2} for a schematic depiction of these interactions, which are all such that the three qubits interact collectively. In the next section we study pairwise interactions.  

In order to characterize the ground state  genuine three-partite entanglement we use the three-tangle (residual entanglement); the pairwise entanglement, via the concurrence of the reduced systems ($12$, $13$, and $23$); and the entanglement of the bipartitions, through the concurrence of the states related to the bipartitions ($1|23$, $12|3$, and $13|2$).

If we take the variance of case (III) to be $\sigma_{123}=\sigma$, then we should have $\sigma_{12}\, \sigma_3=\sigma$ and $\sigma_1\,\sigma_2\,\sigma_3=\sigma$, such that $$\langle \vert { (V_{I})}_{ij}|^2\rangle=\langle \vert (V_{II})_{ij}\vert^2\rangle=\langle \vert (V_{III})_{ij}\vert^2\rangle\propto \sigma.$$
The results only depend on the product of the involved deviations, so for definiteness we adopt the most symmetric choice: $\sigma_{12}=\sigma^{2/3}$ and $\sigma_3=\sigma^{1/3}$ for case (II) and $\sigma_1=\sigma_2=\sigma_3=\sigma^{1/3}$ for case (I). 

In Fig. \ref{fig3}(a) we show the averaged three-tangle for the potentials (I), (II), and (III). The first feature that stands out is the much higher ability of the totally non-separable interactions of type (III) to generate sizable values of $\tau$, with $\langle \tau \rangle$ being more than an order of magnitude larger than that coming from cases (I) and (II). 

In these latter cases, $\langle \tau \rangle$ has a maximum at $\sigma\approx 1$ and decays as $\sigma \to \infty$. This can be understood with the same reasoning as in the previous section: the one-body terms in (\ref{ham3}) become negligible as $\sigma \to \infty$, and the ground states become either separable [case (I)] or bi-separable [case (II)]. The three-tangle, as a genuine three-partite entanglement quantifier, becomes zero in both cases. In case (III), $\langle \tau \rangle$ grows monotonically with $\sigma$ and converges to the average three-tangle of pure, Haar-random three-qubit states, $\langle \tau \rangle_{Haar}=1/3$ \cite{munro}.

The fact that $\langle \tau \rangle$ is very small for cases (I) and (II), at least two orders of magnitude smaller than 1 (the value of $\tau$ for a maximally entangled GHZ state), indicates that the generated ground states are either close to $W$ states or to separable states. We shall return to this point later.

In Fig. \ref{fig3}(b) we show the corresponding standard deviations, which display qualitative behaviors similar to those of the average.

As a complementary characterization, in Fig. \ref{fig4} we plot the average and standard deviation for the total concurrence $ C_t= C_{12}+C_{13}+C_{23}$. Again, it is interaction of type (III) that leads to the highest values. Average total concurrence only decays with $\sigma$ for the totally separable interaction of type (I), as expected. For sizeable values of $\sigma$, the average is twice the standard deviation for type (III) and (II), but half that value for type (I).
\section{Three qubits, pairwise interactions}
\begin{figure}[h]
\includegraphics[height=5cm,angle=0]{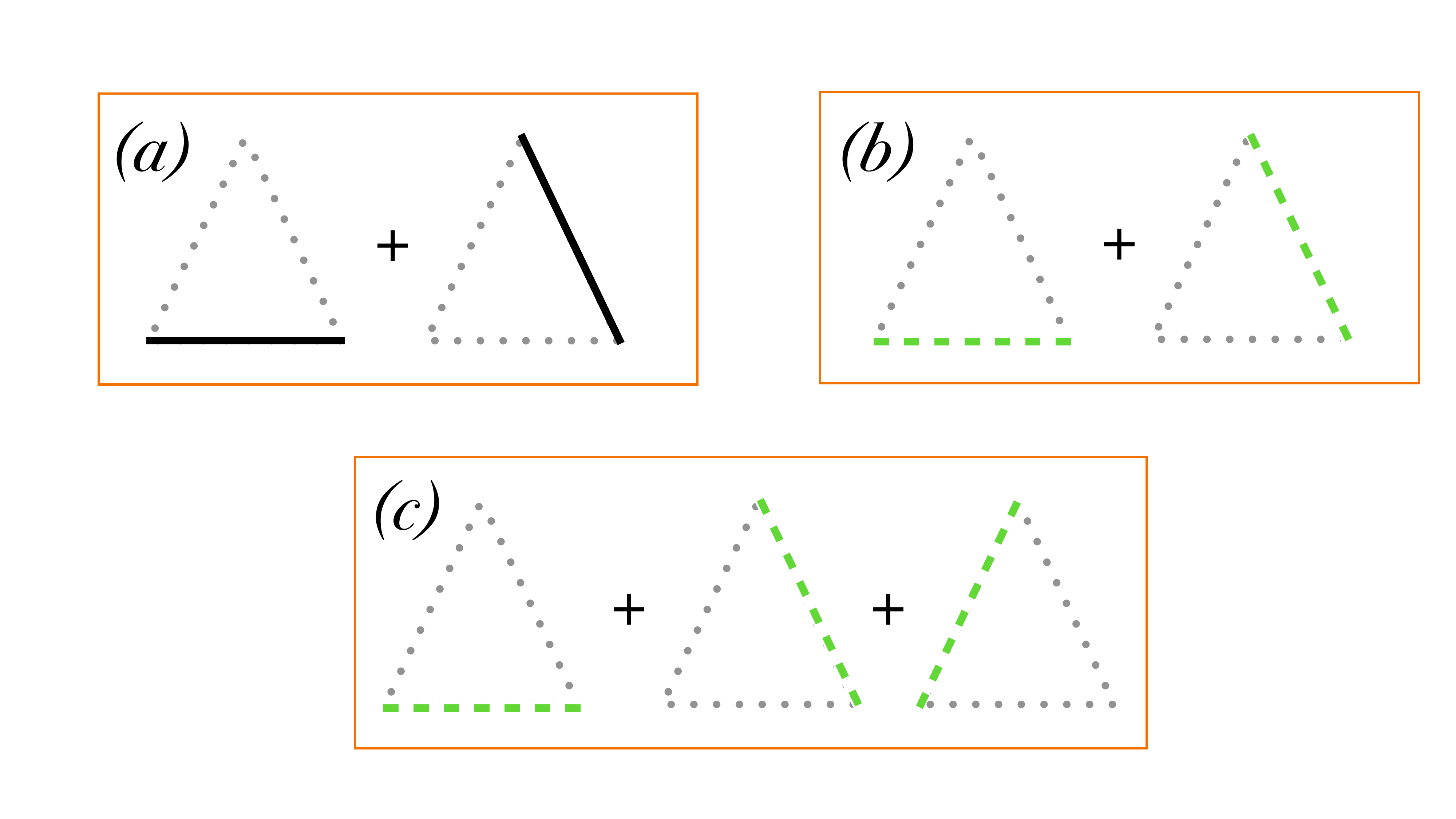}
\caption{Diagrams representing the three types of interactions considered in this section. We use the same convention as in Fig \ref{fig2}. Dotted gray segments denote absence of interaction. (color online)}
 \label{fig5} 
\end{figure}

Here we study random interactions with a pairwise structure which are physically relevant, as for instance:
\begin{eqnarray}
V_a &=& \frac{1}{2} (\mathds{1}_1 \otimes V_{23}+V_{12} \otimes \mathds{1}_3),
\label{rswapp} \\
V_b &=& \frac{1}{2}(\mathds{1}_1\otimes V_{2} \otimes V_3+V_{1} \otimes V_2 \otimes \mathds{1}_3),
\label{rswapp1}
\end{eqnarray}
such that no interaction between qubits 1 and 3 exists. In addition, we consider:
\begin{equation}
V_c=\frac{1}{3}(\mathds{1}_1\otimes V_{2} \otimes V_3+V_{1}\otimes \mathds{1}_2\otimes V_3 +V_{1}\otimes V_2\otimes \mathds{1}_3),
\label{3terms}
\end{equation}
see Fig. \ref{fig5} for a schematic depiction.
The multiplicative factors are there to make the average intensity of the full interaction term the same in all cases. For the same reason, we set $\sigma_{23}=\sigma_{12}=\sigma$ in $V_a$ and $V_b$, and $\sigma_1=\sigma_2=\sigma_3=\sigma^{1/2}$ in $V_c$.

\begin{figure}[ht]
\includegraphics[height=5cm,angle=0]{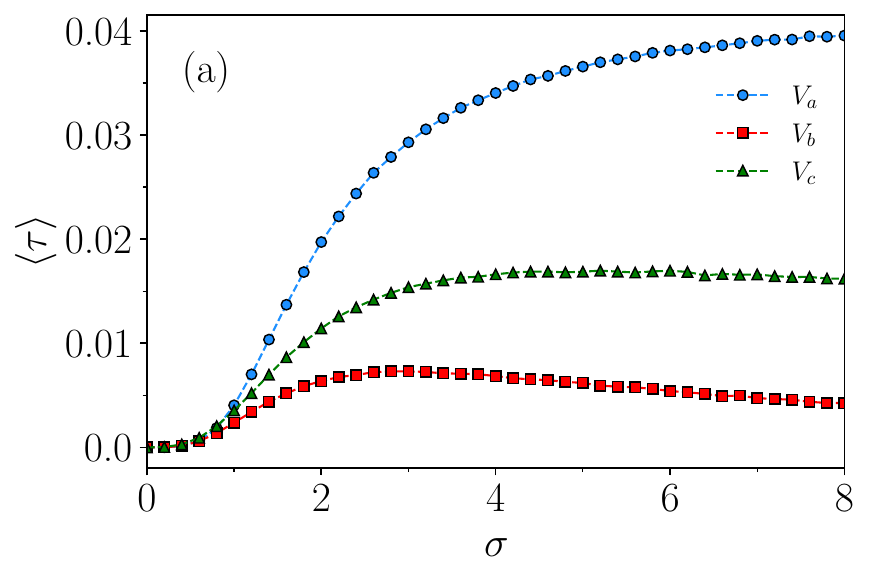}
\includegraphics[height=5cm,angle=0]{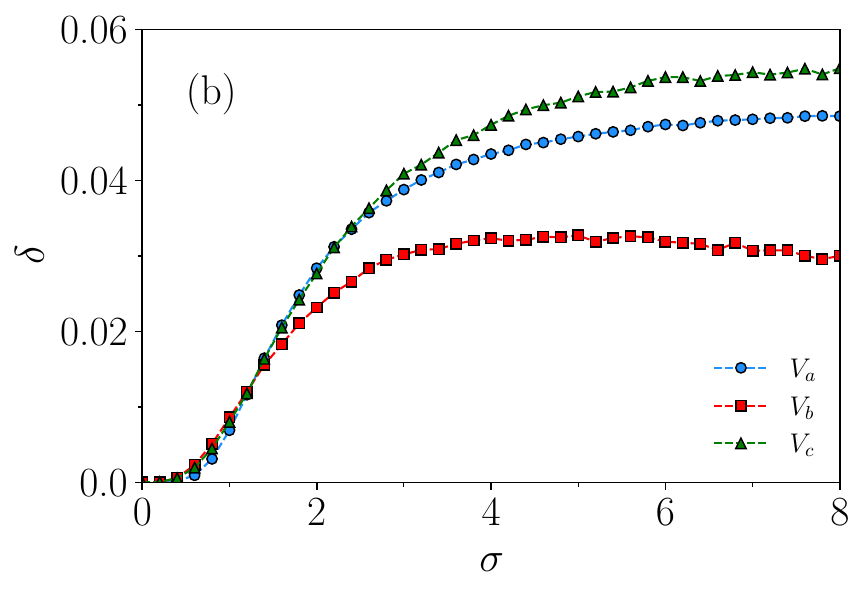}
\caption{(a) Average three-tangle $\left\langle \tau \right\rangle$ from $6 \times 10^5$ samples for the interactions $V_a$, $V_b$, and $V_c$. We find $\left\langle \tau \right\rangle\sim\sigma^3$ for small $\sigma$. (b) Corresponding standard deviation $\delta$. (color online)} 
\label{fig6}
\end{figure}

In figure \ref{fig6} we show the averaged three-tangle of the ground states coming from these interactions, along with the corresponding standard deviations.
Here, for all investigated cases there is a very low degree of genuine three-partite entanglement of GHZ type, as compared to the potential $V_{III}$, for which $\langle \tau \rangle$ is, typically, larger by one order of magnitude [see Fig. \ref{fig4}(a)]. 

A curious feature in Fig. \ref{fig6}(b) is that, although the larger values of $\langle \tau \rangle$ are those produced by $V_a$, the deviations attached to $V_c$ are consistently the larger ones. In all cases the averages and the deviations have the same order of magnitude. 

Notice that in Fig. \ref{fig6}(a), the three-tangle remains close to zero for $0<\sigma<0.5$. In fact, we find that $\langle \tau\rangle\sim\sigma^3$ for small $\sigma$. This will be relevant for future discussions. 

\subsection{Hamiltonian swapping}

The two standard ways to produce entanglement are either to make the parties interact or to carry out a swapping operation (based on entangled measurements) \cite{yurke, zuk}. If, however, we consider the Hamiltonian (\ref{ham3}) with either potential $V_a$ or $V_b$, no interaction occurs between qubits 1 and 3. Notwithstanding, the average ground state entanglement between these qubits is finite (although small). 
\begin{figure}[h]
\includegraphics[height=5cm,angle=0]{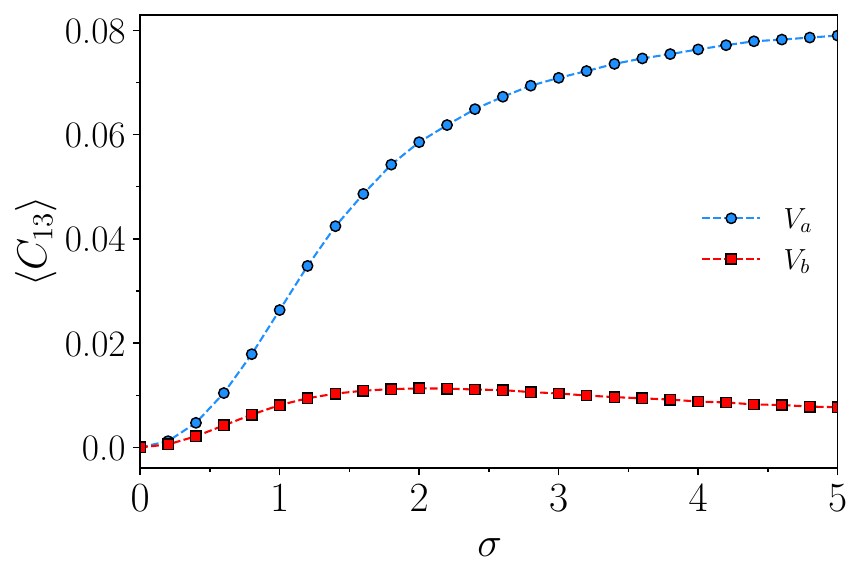}
\caption{Average concurrence $\left\langle C_{13} \right\rangle$ from $5 \times 10^4$ samples for the interactions $V_a$ and $V_b$. (color online)} 
\label{fig7}
\end{figure}
This may be considered as an instance of entanglement swapping without measurement, see for instance \cite{yang}, since neither the correlation appears from a direct interaction nor from an entangling measurement (standard swapping). However, as it can be seen in Fig. \ref{fig7} the average concurrence between 1 and 3 is quite low, saturating around $\langle C_{13} \rangle\sim 0.001$ and $\langle C_{13} \rangle\sim 0.08$ for interactions $V_a$ and $V_b$, respectively. Therefore, the states are close to being bi-separable.

\subsection{Concentration near W states}

As we saw in the previous section, $V_{III}$ gives rise to states with the highest average three-tangle among the investigated interactions. That is to say, this form of $V$ generates GHZ states which are, on average, relatively far from the boundary with W states. The question arises whether there is a form of the random interaction that produces the latter states. The strict answer is negative because that would require fine tuning in order to ensure $\tau=0$. In other words, in the space of parameters of three-qubit pure states ($\chi$), the dimension of the subspace of GHZ states is the same as $\dim \chi$ while, on the other hand, the subspace defined by $\tau=0$ has lower dimension. Therefore, the probability to generate a W state as an eigenvector of a random Hamiltonian is zero. 

This can be understood in a more precise way by employing the optimal parametrization reported in \cite{toni}.
For an arbitrary pure three-qubit state, $\ket{\Psi} $, there is always a basis for which one can write
\begin{equation*}
\ket{\Psi } = a_0 \vert 000 \rangle+a_1 e^{i \varphi} \vert 100 \rangle+a_2 \vert 101 \rangle
+a_3 \vert 110 \rangle+a_4 \vert 111 \rangle,
\end{equation*}
where the coefficients $a_j$ are non-negative real numbers, and $0 \le \varphi \le \pi$. It is easy to show that the corresponding three-tangle is given simply by $\tau=(2a_0 a_4)^2$, so that $\tau=0$ would require either $a_0=0$ or $a_4=0$ (or both). If $a_0=0$ and $a_4\ne 0$ we typically get biseparable states. If $a_0\ne 0$ and $a_4= 0$ we typically obtain W states. Of course, these situations correspond to zero-measure sets, as compared to the set of GHZ states. 

It is important, however, to note that the Haar measure does not correspond to a uniform distribution of the coefficients $a_j$. In particular, it is not correct to state that W and separable states are equally likely. Indeed, for the investigated Hamiltonians, the concurrences concerning any reduced density matrices and bipartitions are typically not zero, even for very low values of three-tangle. 

\begin{figure}[h]
\includegraphics[height=5cm,angle=0]{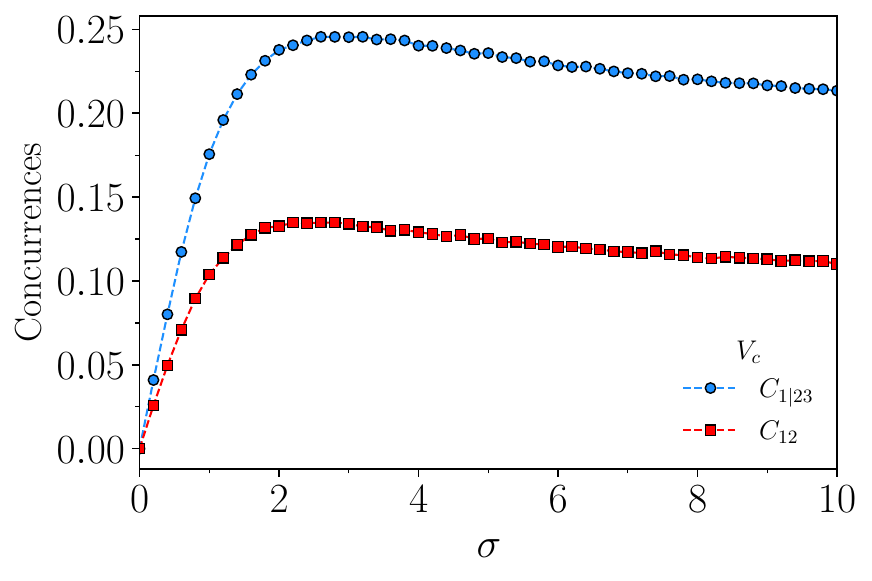}
\caption{$\langle C_{1\vert 23}\rangle$ and $\langle C_{13}\rangle$ with $5 \times 10^4$ samples for the interaction potential $V_c$. Both quantities present a steep linear growth for small $\sigma$. (color online)} \label{conbitpoF}
\label{fig8}
\end{figure}

Consider, for instance, the symmetric interaction potential $V_c$. It leads to ground states with $\tau<0.05$ for any value of $\sigma$, as one can see in Fig. \ref{fig6}(a). On the other hand, the concurrences of bipartitions and reduced systems are sizable as can be seen in Fig. \ref{fig8}, where $C_{12}=C_{13}=C_{23}$ and $C_{1|23}=C_{2|13}=C_{3|12}$, due to the mentioned symmetry. This indicates a possible close proximity to W states.

For small values of $\sigma$ we found $C_{12}\sim \sigma$ and $C_{1|23}\sim \sigma$. From Eq. (\ref{tau}), one might expect $\tau \sim \sigma^{2}$. However, the second order terms cancel out and we obtain $\tau \sim \sigma^{3}$. That is to say, in this region we may have very small three-tangle and not so small concurrences.

\begin{figure}
\includegraphics[scale=0.5]{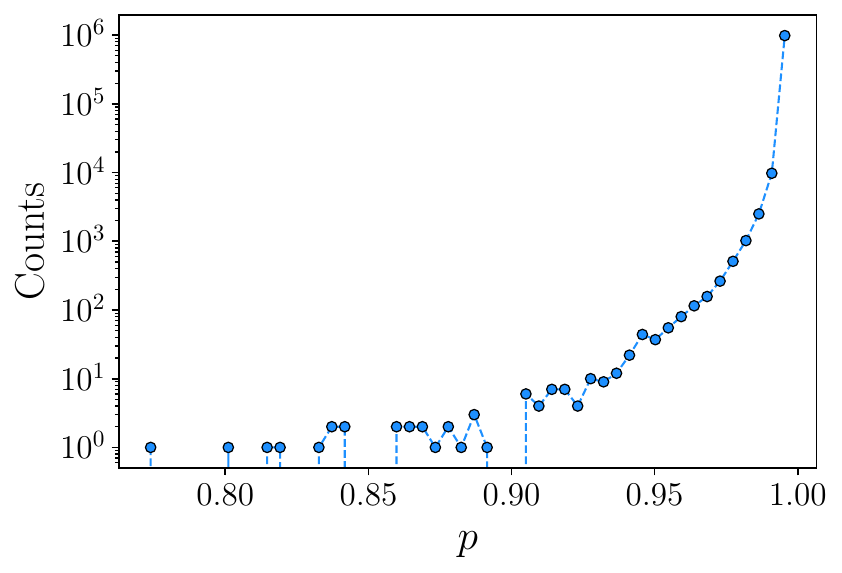}
\caption{Overlap $p=\vert \langle \Psi \vert \Phi_N \rangle \vert^2$ between the Hamiltonian's ground state and the nearest $\tau=0$ state, for $\sigma = 0.5$. From $10^6$ samples.}
\label{fig9}
\end{figure}

For a more quantitative analysis, let us consider the ground state of $V_c$ with $\sigma=0.5$. We computed $10^6$ ground states $|\Psi\rangle$, and computed the overlap of each with the nearest state having $\tau=0$ (numerically $\tau<10^{-6}$), which we denote $|\Phi_N\rangle$. We get $99.89 \%$ of ground states with $p=\vert \langle \Psi \vert \Phi_N \rangle \vert^2>0.98$, see Fig. \ref{fig9}.
For these states, $C_{12}<0.4$, with about $14 \%$ of the states with $C_{12} > 0.1$. In addition, 
we found that $\tau<{\rm min}\{C_{12}^2,C_{13}^2,C_{23}^2\}$ for more than $99.99\%$ of the states $|\Phi_N\rangle$.
Usually, the three-tangle is several orders of magnitude smaller than the smallest two-tangle.

%

We conclude that the ground states are very close to being W states, and not so close to being separable. So, in practice, the GHZ states so generated are almost indistinguishable from the nearest W state. Although the formal difference between these two classes of states is well defined, we may find actual situations for which the actual numerical and practical distinction become difficult.

\section{Closing remarks}
In this work we investigated the entanglement in the ground state of Hamiltonians containing deterministic one-body terms (spins subjected to a specified magnetic field) and random interaction terms, for two and three qubits. By varying the relative intensity or these contributions one can interpolate between separable and fully random (Haar distributed) states. It is clear that, although the characterization of Haar random states is relevant, restricting attention to them leaves a wealth of physically relevant situations unaddressed.

We found that the amount and nature of the resulting entanglement strongly depends on the underlying topology of the interaction terms. We considered three types of collective interaction and three types of pairwise interaction, all differing in their degrees of separability. For three qubits we found strong GHZ entanglement with a fully non-separable collective interaction (a random matrix from $GUE(8)$),  and the production of near-W states for the opposite case of a fully separable pairwise interaction.

An interesting perspective is to increase the number of qubits in order to investigate more deeply the role of interaction topology in entanglement production. Where should we generically expect to find more multipartite entanglement, in the ground state of a fully connected network of spins, or in a system with many pairwise interactions? The answer probably depends on the type of entanglement that is required. 

\begin{acknowledgments}

This work received financial support from the Brazilian agencies Coordena\c{c}\~ao de Aperfei\c{c}oamento de Pessoal de N\'{\i}vel Superior (CAPES), Funda\c{c}\~ao de Amparo \`a Ci\^encia e Tecnologia do Estado de Pernambuco (FACEPE), Conselho Nacional de Desenvolvimento Cient\'{\i}fico  e Tecnol\'ogico (CNPq) and Funda\c{c}\~ao de Amparo do Estado de Goi\'as (FAPEG). FP acknowledges the support from the INCT-IQ  program (Grant 465469/2014-0), and Funda\c{c}\~ao de Amparo \`a Pesquisa do Estado de S\~ao Paulo (FAPESP - Grant 2021/06535-0). PFG acknowledges the support of FAPEG and CNPq (grant 405508/2021-2).
\end{acknowledgments}

\end{document}